\documentclass[sn-mathphys-num]{sn-jnl}

\usepackage{graphicx}%
\usepackage{multirow}%
\usepackage{amsmath,amssymb,amsfonts}%
\usepackage{amsthm}%
\usepackage{mathrsfs}%
\usepackage[title]{appendix}%
\usepackage{xcolor}%
\usepackage{textcomp}%
\usepackage{manyfoot}%
\usepackage{booktabs}%
\usepackage{algorithm}%
\usepackage{algorithmicx}%
\usepackage{algpseudocode}%
\usepackage{listings}%

\theoremstyle{thmstyleone}%
\theoremstyle{thmstyletwo}%

\theoremstyle{thmstylethree}%

\newcommand{\ket}[1]{|#1\rangle}
\newcommand{\bra}[1]{\langle #1|}

\begin{document}

\title[Boundary effect and quantum phases in spin chains]{Boundary effect and quantum phases in spin chains}


\author[1]{\fnm{Jinhyeok} \sur{Ryu}}\email{jisungin28@gmail.com}
\author*[1]{\fnm{Jaeyoon} \sur{Cho}}\email{choooir@gmail.com}

\affil*[1]{\orgdiv{Department of Physics and Research Institute of Natural Science}, \orgname{Gyeongsang National University}, \orgaddress{\city{Jinju}, \postcode{52828}, \country{Korea}}}


\abstract{
    Boundary effect is a widespread idea in many-body theories. However, it is more of a conceptual notion than a rigorously defined physical quantity. One can quantify the boundary effect by comparing two ground states of the same physical model, which differ only slightly in system size. Here, we analyze the quantity, which we call a boundary effect function, for an $XXZ$ spin-$\frac{1}{2}$ model using density matrix renormalization group calculations. We find that three quantum phases of the model manifest as different functional forms of the boundary effect function. As a result, the quantum phase transition of the model is associated with a nonanalytic change of the boundary effect function. This work thus provides and concretizes a novel perspective on the relationship between bulk and boundary properties of ground states.
}




\maketitle

\section{Introduction}

One of the primary subjects in quantum many-body theory is to understand the characteristics of many-body ground states.
A particularly interesting scenario occurs when the Hamiltonian contains mutually competing terms.
When their ratio varies, the ground state typically undergoes a (second-order) quantum phase transition.
The Landau-Ginzburg-Wilson (LGW) paradigm asserts that a local order parameter, associated with the intrinsic symmetry of the system, signifies the quantum phases, and the transition accompanies the nonanalyticities of the fundamental scales, constituting critical phenomena~\cite{wil74,son97,sac11}.
For example, the characteristic length and time (or inverse energy) scales diverge at the critical point, accounting for the sudden, global change of the internal structure.

The quantum phases and phase transitions essentially require a thermodynamic limit to introduce strict nonanalyticities of physical quantities.
The existence of a \emph{local} order parameter implies that ground states with different quantum phases look different locally.
While this appears to be a necessary premise in classifying translation-invariant infinite-sized ground states, there exist various quantum matters defying this concept.
For example, topological orders are attributed to ground states that are locally indistinguishable, yet have different nonlocal%
\footnote{Here, ``nonlocal'' usually means a length scale much larger than the correlation length.} 
or dynamical characteristics, such as topological entanglement entropies~\cite{kit06,lev06}, fractional quasi-particle statistics~\cite{wil82,aro84}, and topological degeneracies~\cite{wen90}.
In particular, when such a system is contained in a bounded space, robust edge modes appear at the boundary~\cite{wen90b,wen91}.

In topologically ordered phases, the edge mode reflects the topological number defined as a bulk property.
This notion of bulk-boundary correspondence has attracted a good deal of interest in modern condensed matter physics~\cite{wen91,hat93,lu12}.
However, there is no necessary reason to believe that such a notion belongs exclusively to topological phases.
In any form of quantum phase, including those within the LGW framework, the idea that bulk properties manifest somehow at the boundary seems plausible.
Of course, the materialization of this idea is nontrivial.
Unlike the topological phases wherein the boundary modes appear distinctively, the boundary effect in the LGW framework would merely alter the already existing ground state.
A naturally arising question is then ``which properties manifest at the boundary in what form?''.

\begin{figure}
    \centering
    \includegraphics[width=0.7\textwidth]{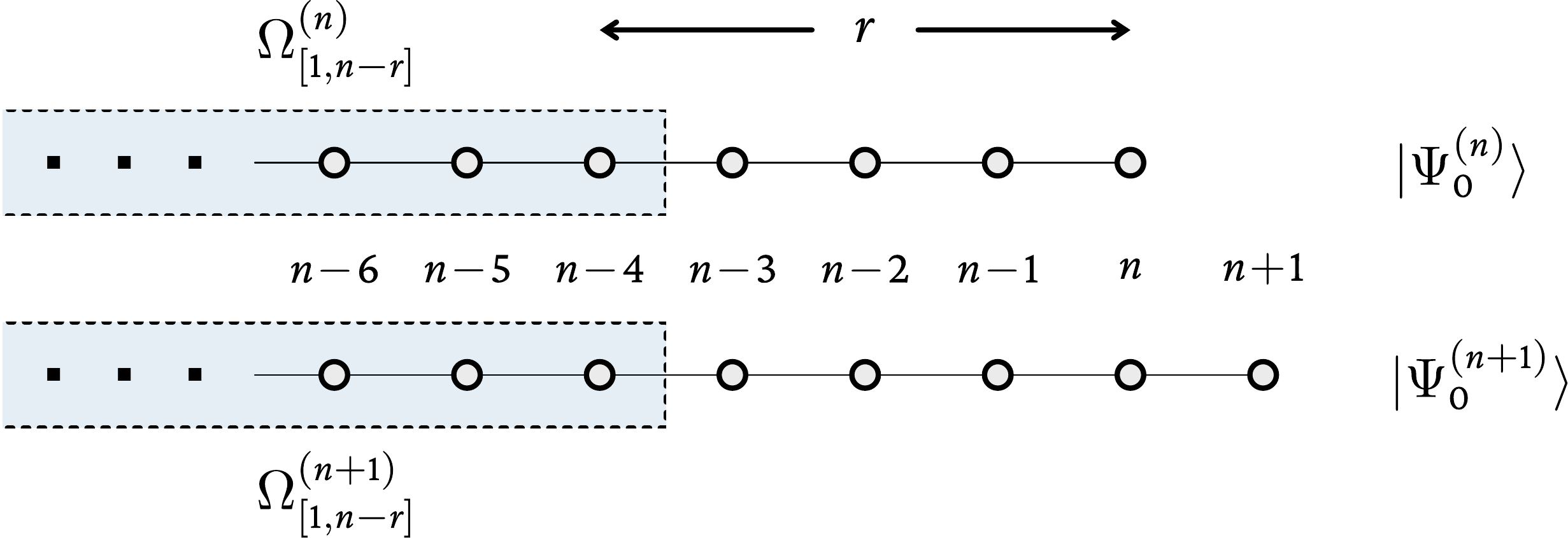}
    \caption{To obtain boundary effect function $\mu_n(r)$, the ground states $|\Psi_0^{(n)}\rangle$ and $|\Psi_0^{(n+1)}\rangle$ of the same spin model with different system sizes $n$ and $n+1$ are compared. The two reduced density matrices $\Omega_{[1,n-r]}^{(n)}$ and $\Omega_{[1,n-r]}^{(n+1)}$ for the sites in the shaded boxes become closer as $r$ increases, decreasing $\mu_n(r)$.}
    \label{fig1}
\end{figure}

In this context, the connection between the correlation length in the bulk and the behavior of the boundary effect as defined in Refs.~\cite{cho14,cho15} provides an intriguing perspective.
The idea is illustrated in Fig.~\ref{fig1} for one-dimensional cases.
Consider a particular spin model with the Hamiltonian determined for any system size in a consistent way (for example, imagine Ising chains, Heisenberg chains, etc.).
Let $H^{(n)}$ be the Hamiltonian when the system size (or the number of spins) is $n$ and $|\Psi_0^{(n)}\rangle$ be the ground state of $H^{(n)}$.
As our concern is the effect of the boundary, we consider an open boundary condition.
As in Fig.~\ref{fig1}, we take the reduced density matrix of $|\Psi_0^{(n)}\rangle$ by removing $r$ spins near the boundary, which we denote by $\Omega_{[1,n-r]}^{(n)}$.
Similarly, we take the reduced density matrix $\Omega_{[1,n-r]}^{(n+1)}$ for the same sites from a slightly larger ground state $|\Psi_0^{(n+1)}\rangle$.
Denoting by $F(\rho,\sigma) \equiv \text{Tr} \sqrt{\sqrt{\rho} \sigma \sqrt{\rho}}$ the fidelity between two density matrices $\rho$ and $\sigma$, we define what we call a boundary effect function (BEF) as
\begin{equation}
    \mu_n(r) = \sqrt{1 - F \left(\Omega_{[1,n-r]}^{(n)}, \Omega_{[1,n-r]}^{(n+1)} \right)}.
    \label{eq:bef}
\end{equation}
By definition, $\mu_n(r)$ is a non-increasing function of $r$.
For sufficiently large $n$ and $r$, the two density matrices $\Omega_{[1,n-r]}^{(n)}$ and $\Omega_{[1,n-r]}^{(n+1)}$ would be indistinguishable as the state in the deep bulk region (the shaded region in Fig.~\ref{fig1}) would converge to the state in the thermodynamic limit.
This implies that $\mu_n(r) \rightarrow 0$ as $r \rightarrow \infty$.
In this sense, the BEF quantifies how far the boundary effect spreads to the direction toward the bulk.

Interestingly, the behavior of the BEF restricts the nature of correlation in the bulk~\cite{cho15}.
It can be shown that if $\mu_{\infty}(r)$ decays exponentially with $r$, the bulk has a finite correlation length, called the exponential clustering~\cite{fre85,has06,nac06}, and exhibits an area-law scaling of entanglement entropies~\cite{eis10}.
Consequently, in addition to the known interplay between the spectral gap, correlation length, and entanglement, the BEF is also intimately related to them.
These relations result from a different viewpoint, where a large-sized ground state is regarded as a state \emph{grown up} from a small one; during this process, correlation is developed as a trail of the expansion.
Note, however, that this picture leaves the BEF only as an \emph{upper} bound to the correlation.
It is unknown if a finite correlation length, i.e., exponential decay of all correlation functions in the bulk, in turn implies an exponentially decaying BEF.
In fact, a proof of the tightness of the upper bound, if possible, leads to a proof of the entanglement area law in any dimension, producing far-reaching consequences to diverse fields in theoretical physics~\cite{cho15,eis10}. 
It is thus worthwhile to investigate the nature of BEFs for various physical models.
This investigation would also concretize the aforementioned idea that bulk properties can be manifested at the boundary in ordinary matters.

The aim of this paper is to analyze the BEF for an $XXZ$ spin-$\frac{1}{2}$ chain model, which has three distinct quantum phases~\cite{gia03}.
Our analysis is aided by density matrix renormalization group (DMRG) calculations~\cite{sch11}.
We find that the functional form of the BEF reflects the quantum phase in the bulk.
Consequently, the qualitative change of the BEF signifies the quantum phase transition.
Our finding thus adds yet another attribute to the list of properties associated with quantum phases and phase transitions.
Our finding also indicates that for the XXZ spin-$\frac{1}{2}$ chain, the upper bound set by the BEF is a \emph{tight} one.
This concretizes the aforementioned picture regarding the interplay between the BEF and various correlation characteristics of the ground state.

\section{Boundary effect functions for matrix product states}

\begin{figure}
    \centering
    \includegraphics[width=0.76\textwidth]{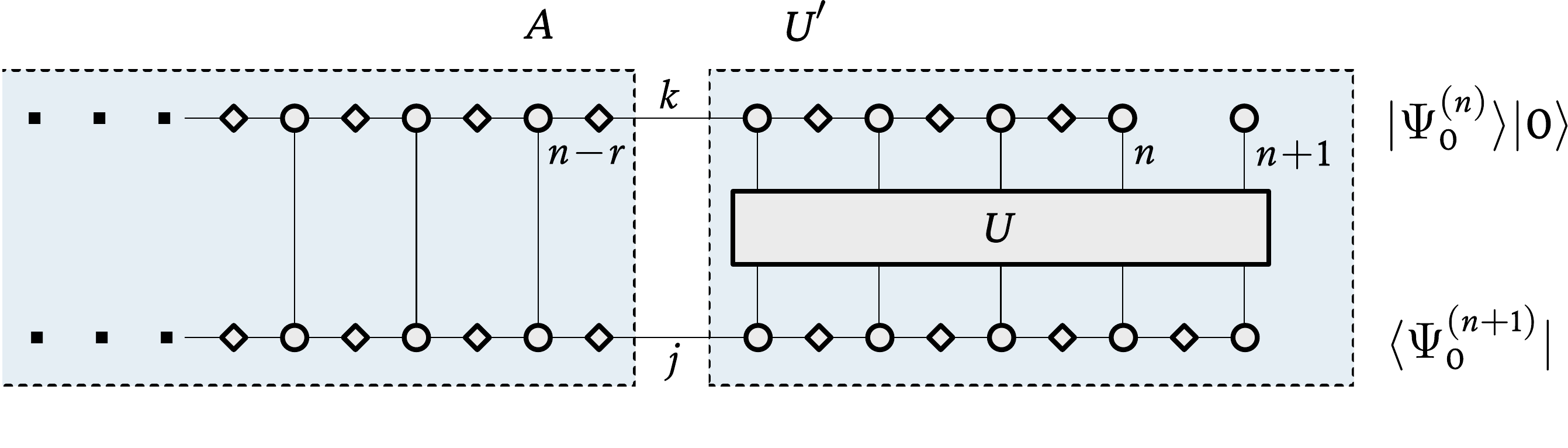}
    \caption{The tensor network diagram for the calculation of the Uhlmann fidelity between two reduced density matrices of $\ket{\Psi_0^{(n)}}$ and $\ket{\Psi_0^{(n+1)}}$. The absolute value of the result is taken and maximized over the variable unitary matrix $U$. The fidelity is then identical to the trace norm of the matrix $A$ on the left-hand side.}
    \label{fig2}
\end{figure}

The BEF~\eqref{eq:bef} contains the fidelity between reduced density matrices.
For matrix product states, the calculation of the fidelity is best illustrated by the conventional diagrammatic representation~\cite{sch11,hau18}.
Using the Uhlmann fidelity~\cite{wil13}, we have
\begin{equation}
    F\left(\Omega_{[1,n-r]}^{(n)}, \Omega_{[1,n-r]}^{(n+1)}\right) 
    = 
    \max_{U}\left|\bra{\Psi_0^{(n+1)}} I_{[1,n-r]} \otimes U_{[n-r+1,n+1]} \ket{\Psi_0^{(n)}}\ket{0}\right|,
\end{equation}
where $\ket{0}$ is an arbitrary state at site $n+1$ and $U$ ($I$) is a unitary (identity) matrix acting on the sites specified by the subscript.
This can be represented by the diagram in Fig.~\ref{fig2}, where it is understood that the absolute value of the result is taken and maximized over the variable unitary matrix $U$.
Here, both states are represented in a mixed-canonical form~\cite{sch11}.
Let us divide the diagram into two parts as in the figure.
The upper tensor on the right-hand side represents the orthonormal bases of that part when $\ket{\Psi_0^{(n)}}\ket{0}$ is Schmidt decomposed.
Similarly, the lower tensor represents the orthonormal bases for $\bra{\Psi_0^{(n+1)}}$.
Combined with $U$, the right-hand side as a whole thus represents a variable unitary matrix, which we denote by $U'$.
The tensor $A_{jk}$ on the left-hand side in the figure is a matrix and can be singular-value decomposed as $A = V S W$ with $V$ and $W$ being unitary and $S$ being positive real diagonal.
We are led to
\begin{equation}
    \begin{split}
        F\left(\Omega_{[1,n-r]}^{(n)}, \Omega_{[1,n-r]}^{(n+1)}\right) 
        &= \max_{U'} \sum_{j, k} (VSW)_{jk}U'_{kj} = \max_{U'} \sum_{j} S_{jj} (WU'V)_{jj} \\
        &= \sum_j S_{jj} = \| A \|_1,
    \end{split}
\end{equation}
where the last result denotes the trace norm of $A$.
Consequently, the problem reduces to obtaining the matrix $A$ by contracting the relevant parts of the matrix product states.

\section{Boundary effect functions in $XXZ$ spin chains}

We examine the BEF for $XXZ$ spin-$\frac{1}{2}$ chains governed by Hamiltonian
\begin{equation}
    H = \sum_{j=1}^{n-1} \left(\sigma_j^x \sigma_{j+1}^x + \sigma_j^y \sigma_{j+1}^y + \Delta \, \sigma_j^z \sigma_{j+1}^z \right),
\end{equation}
where $\sigma_j^x$, $\sigma_j^y$, and $\sigma_j^z$ are the Pauli operators acting on the $j$-th spin.
This model has three distinct quantum phases: the gapped Ising ferromagnet for $\Delta<-1$, the gapless $XY$ phase for $|\Delta| < 1$, and the gapped Ising anti-ferromagnet for $\Delta>1$~\cite{gia03}.
For the former two phases, the nature of the BEF is predictable, as explained below.
The main focus of our numerical analysis is thus concerning the Ising anti-ferromagnetic phase.

\subsection{$\Delta<1$: Ising ferromagnetic phase}

The ground state problem for $\Delta<1$ is trivial.
For any $\Delta<1$, the ground state has a two-fold degeneracy and is spanned by $\ket{\uparrow\uparrow\cdots}$ and $\ket{\downarrow\downarrow\cdots}$, where $\ket{\uparrow}$ and $\ket{\downarrow}$ are the eigenstates of $\sigma^z$.
Once the system falls into one of the two exact ferromagnetic states by spontaneous symmetry breaking, the ground state has no correlation at all between different sites.
As a result, the BEF $\mu_n(r)$ vanishes for any $n$ and $r$, implying that the size of the ground state can be grown by simply adding separate spins at the boundary.

\subsection{$-1<\Delta<1$: $XY$ phase}

For $|\Delta|<1$, the correlation length of the ground state diverges in the thermodynamic limit.
In such cases, it can be shown that the BEF $\mu_n(r)$ can not decay exponentially with $r$~\cite{cho15}.
As a slowly decaying BEF does not provide a practically useful upper bound to correlations and entanglement in the ground state, the precise behavior of $\mu_n(r)$ is relatively unimportant in analyzing many-body ground states.
Apart from that, it is obvious that in crossing the critical point $\Delta=-1$, the BEF should undergo a sudden transition in nature.

In this phase, the DMRG calculation produces essentially unreliable results as the required dimension of the bonds in matrix product states increases with the system size.
This is particularly problematic in our analysis, where states are compared site by site.
For this reason, we omit the results of our DMRG calculation for this phase.
The results are similar to those for small $\Delta$ in Subsection~\ref{subsec:antiferro}.

\subsection{$\Delta>1$: Ising anti-ferromagnetic phase\label{subsec:antiferro}}

\begin{figure}
    \centering
    \includegraphics[width=0.4\textwidth]{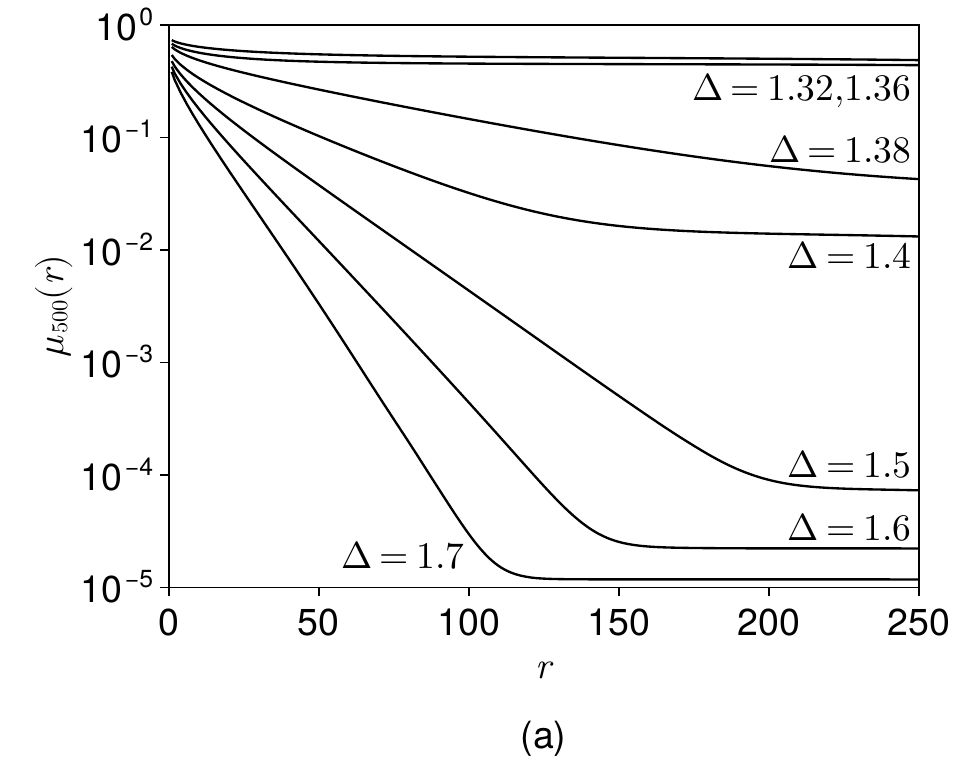}
    \includegraphics[width=0.4\textwidth]{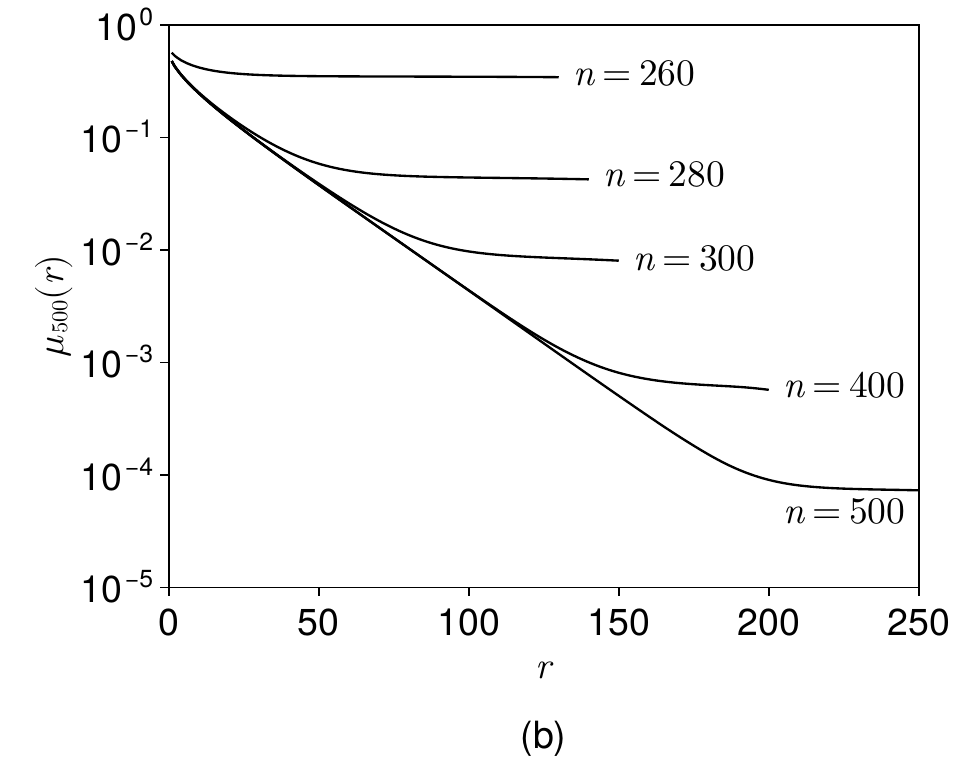} 
    \caption{Boundary effect functions for the Ising anti-ferromagnetic phase ($\Delta>1$). In (a), $\mu_{500}(r)$ is plotted for various values of $\Delta$. To examine the finite-size effect, $\mu_r(r)$ is plotted in (b) for various $n$ while fixing $\Delta=1.5$.}
    \label{fig3}
\end{figure}

For $\Delta>1$, the ground state has a finite spectral gap to the excited state, hence a finite correlation length.
Our primary concern is whether the BEF $\mu_n(r)$ also decays exponentially with $r$ for large $n$.

We have performed DMRG calculations for this phase using the ITensor software library~\cite{itensor}.
In obtaining the ground state, we ensured the convergence of the state by repeating the DMRG procedure until a single sweep reduces the energy with a rate less than $10^{-8}$.
We have chosen the parameters such that the maximum bond dimension of matrix product states is 200 and the truncation error cutoff is $10^{-12}$.
In fact, the choice of the simulation parameters is not crucial in recognizing the essential characteristics of the BEF in this phase, as will be explained below.
The ground state has a two-fold degeneracy as the Hamiltonian is invariant under flipping all $\sigma^z$ spins.
We have handled this issue by choosing states with the same aligning of $\sigma_1^z$ of the first spin.
While this strategy makes sense in accordance with the picture of spontaneous symmetry breaking, we note that the essential behavior of the BEF does not change even when the two oppositely ordered ground states are superposed.
We obtain $\mu_n(r)$ only for $r<n/2$ because for larger $r$, the effect from the opposite boundary (site $1$) would be stronger.

In Fig.~\ref{fig3}(a), we plot the BEF $\mu_{500}(r)$ for various $\Delta$.
Let us start from the result for $\Delta=1.7$ and examine how it changes when $\Delta$ decreases, approaching the critical point $\Delta=1$.
For $\Delta=1.7$, $\mu_{500}(r)$ shows a remarkably clean signature of exponential decay with respect to $r$.
Moreover, it saturates sharply at around $r_s=110$.
For convenience, let us call this $r_s$ the \emph{saturation length}.
The saturation appears to originate from the interference of the two boundary effects from the opposite sides.
This is elaborated below with the result in Fig.~\ref{fig3}(b).

The saturation length varies from time to time at each instance of the simulation.
Nonetheless, the initial slope of the exponential decay is almost invariant.
We have also varied the tuning parameters of DMRG and yet the slope is invariant despite the change of the saturation length.
This supports the assertion that the exponential decay of the BEF is an intrinsic nature of this phase.

Fig.~\ref{fig3}(a) indicates that as $\Delta$ decreases, or the critical point $\Delta=1$ is approached, the exponential decay of $\mu_n(r)$ becomes slower.
This is an anticipated result as BEFs are lower bound by correlation functions.
Notable features begin to appear when the saturation length exceeds half of the chain length.
Before getting to this point, it helps to examine the result from a different perspective.

In Fig.~\ref{fig3}(b), we plot $\mu_n(r)$ with different chain length $n$ for $\Delta=1.5$ fixed.
The exponential decay of $\mu_n(r)$ is already evident for $\Delta=1.5$ from Fig.~\ref{fig3}(a).
It can be seen that while the initial decay of $\mu_n(r)$ is independent of $n$, the saturation length decreases with decreasing $n$.
A reasonable interpretation is that this is ascribable to the interplay of the two opposite boundary effects, which is a finite-size effect.
The saturation length almost vanishes around $n\sim 260$, for which the boundary effects overwhelm the entire chain.
In short, the ratio of the saturation length to the chain length is determined by the slope of $\log[\mu_n(r/n)]$.

Returning to Fig.~\ref{fig3}(a), we can understand the behavior for small $\Delta$ from the analysis in Fig.~\ref{fig3}(b).
When $\Delta$ decreases, the decay rate of the BEF also decreases.
When this rate is too slow with respect to the chain length, the initial exponential decay of the BEF is not observable, as elaborated above.
In the thermodynamic limit, such a finite-size effect is absent and the exponential decay of the BEF should be observable for any $\Delta>1$.

\section{Conclusion}

In this work, we have investigated the BEF for various parametric regimes of the $XXZ$ spin-$\frac{1}{2}$ chain.
The model has three quantum phases, namely, an Ising ferromagnetic phase, an $XY$ phase, and an Ising antiferromagnetic phase.
In the former two cases, the behavior of the BEF is analytically predictable.
In the Ising ferromagnetic phase, which is gapped, the ground state has a strict ferromagnetic order throughout the entire parameter space, for which the BEF is trivially determined to vanish.
In the $XY$ phase, which is gapless, the diverging correlation length in the bulk prohibits a superpolynomial decay of the BEF. 
Consequently, the transition between these two phases is associated with a sudden change of the characteristics of BEFs.
The most difficult part to analyze is the Ising antiferromagnetic phase, which requires a numerical treatment.
We have investigated the BEF for this gapped phase through the DMRG calculation.
We found that BEF $\mu_n(r)$ decays exponentially with $r$ in this phase.
In addition, we unfolded the finite-size effect originating from an interplay of two boundary effects from the opposite sides, leading to a conclusion that the exponential decay of the BEF is an intrinsic characteristic of the Ising antiferromagnetic phase.

We would like to emphasize that while a correlation length is a property of a single state, the BEF is one extracted from comparing two ground states.
Consequently, there is no guarantee that a finite correlation length of the bulk leads to an exponential decay of BEFs.
Rather, it is a \emph{desirable} property yet to be established rigorously.
If it turns out to be the case, one can prove one of the long-standing open problems on many-body entanglement, producing widespread impacts~\cite{cho15,eis10}.
In this sense, our finding is an important step in quantum information approaches to many-body theory.

\bmhead{Acknowledgements}

This work was supported by the National Research Foundation (NRF) of Korea under Grant No.~NRF-2022R1A4A1030660.


\end{document}